# Procedures for proper validation of record critical current density claims

Chiara Tarantini and David C. Larbalestier

National High Magnetic Field Laboratory, Florida State University, Tallahassee, FL 32310, USA

E-mail: tarantini@asc.magnet.fsu.edu

**Abstract**

High critical current density $J_c$ is generally the most important property of any superconductor used for applications. The critical current $I_c$ is most frequently measured by transport 4 probe V-I techniques and the measured $I_c$ is then divided by the cross-sectional area to give the critical current density $J_c$. This method is normally error-free. Occasionally however, $I_c$ and $J_c$ are computed from measurements of the magnetic moment and two recent papers (Goyal *et al.* 2024 Nat. Commun. 15, 6523 and Polichetti *et al.* 2024 arXiv:2410.09197) illustrate some of the perils of reporting properties of $RE$Ba$_2$Cu$_3$O$_{7-\delta}$ ($RE$BCO) films grown on a coated conductor substrate using only magnetic evaluations. Correct data analysis is made difficult by the persistent use of cgs units in commercial instrument outputs and by misapplication of handbook Bean model formulae, which in this case led to an order of magnitude $J_c$ overestimate, which of course renders the original world record claim invalid. The laudable attempt of the authors to explain their original mistake has unfortunately its own errors. Here we explain explicitly how errors are still present in the second paper and provide correct expressions for the conversion of magnetic moments to $J_c$ values using Bean model expressions. An important conclusion is that record $J_c$ values ought always to be validated by real transport measurements and by cross-reference to similar samples and their reported $J_c$ values.

**Introduction**

Recently an article by Goyal *et al*. [1] has attracted much attention because of its claim of world record critical current density values in $RE$Ba$_2$Cu$_3$O$_{7-\delta}$ ($RE$BCO, $RE$ being rare earths) film samples. However, the values were obtained only by magnetic moment measurements, from which $J_{c,mag}$ was calculated using unstated expressions. In oral presentations of the results, it was several times suggested that the values were so high relative to the depairing current density that transport measurement verification was advisable but when the paper was published on August 7[th], 2024, only magnetic characterizations supported their claim of world-record $J_c$ values. Talantsev and Tallon [2] asked the authors for the raw data and recalculated $J_{c,mag}$ using SI formula and found that the actual $J_{c,mag}$ values were 10 times smaller than those reported [1] because of a unit conversion error. Once the error was revealed, the authors withdrew the paper (officially retracted on October 23[rd], 2024) [3]. In a follow up report, Polichetti *et al*.[4] explained how the error occurred and provided their own formulae as a guide to others of how to avoid similar error in the future.

In their follow up paper, Polichetti *et al*. [4] suggest the following expressions for correct $J_{c,mag}$ calculation:

*"To summarize:*

$$\frac{2\times \Delta M[A/m]}{w[m]\times\left(1-\frac{w[m]}{3b[m]}\right)} = J_{c,mag}[A/m^2] \qquad (3)$$

*or*

$$\frac{2\times\Delta M[emu/cm^3]}{w[cm]\times\left(1-\frac{w[cm]}{3b[cm]}\right)} = \frac{20\times\Delta M[A/cm]}{w[cm]\times\left(1-\frac{w[cm]}{3b[cm]}\right)} = J_{c,mag}[A/cm^2] \qquad (4)\text{" from ref. [4]}$$

However, equations 3 and 4 are inconsistent. Whereas equation (3) in ref. [4] is correct and in SI units, eq. (4) is incorrect in both the first and the second terms, as we show below. Indeed, as noted by Polichetti [4], a problem is that literature formulae are frequently expressed in different unit systems, not always specified, making unit conversion uncertain.

Because of our own experience in the difficulties of changing units and to avoid further proliferation of incorrect formulae, we decided to test the agreement of Polichetti's eqs. (3) and (4), working from first principles. First, we calculate $J_{c,mag}$ in SI units for a sample of rectangular cross-section in perpendicular field using the simplest Bean model [5], [6] scenario, being explicit about how to properly convert the quantities into SI units. We then compare this formula against eqs. (3) and (4) in [4]. Then, since many still use cgs-emu units because of the way that the most popular magnetometers present their output, we present a converted formula allowing direct use of cgs-emu quantities to $J_{c,mag}$. This then reveals the errors in Polichetti et al.'s eq. (4) [4].

An important point of ref. [4] is to explain how easy it is to make such errors and it is thus absolutely pertinent to cite the final paragraph of Goldfarb's paper [7], which earlier explored this territory too. He cites and comments on a more than 100 year old paper by Giorgi [8]: "In introducing his rationalized, four-dimensional system, Giorgi [1901] wrote, *Il sistema CGS, con questo, perde ogni ragione di esistere; ma non credo che il suo abbandono sarà lamentato da alcuno*. ("With this, the CGS system loses every reason to exist; but I do not think that its abandonment will be lamented by anyone."). He may be correct, eventually." [7].

**Bean model calculation of $J_{c,mag}$ in SI units and practical conversion**

In SI units the moment generated by a current circulating in a loop is given by $m = IA_{loop}$, where $I$ is the current in [A] and $A_{loop}$ is the area of the loop in [m²] (notice that this equation is not valid in Gaussian units where $m = IA_{loop}/c$; in fact Gaussian and cgs-emu units are the same only for magnetic properties as stated clearly in [9]). Assuming the simplest Bean scenario in which the field B has fully penetrated the film, the induced current in a sample of rectangular cross-section circulates in rectangular loops with the same absolute value of $J_{c,mag}$ at all points in the cross-section, as shown in Figure 1, and with no changes through the sample thickness $t$. Here we assume that the sample cross-section is $w \times l = 2a \times 2b$, with $a < b$. Considering the sketch in Figure 1 and its symmetry, the differential magnetic moment can be expressed as:

$$dm = dIA_{loop} = (J_c t dx)(4(b - a + x)x)$$

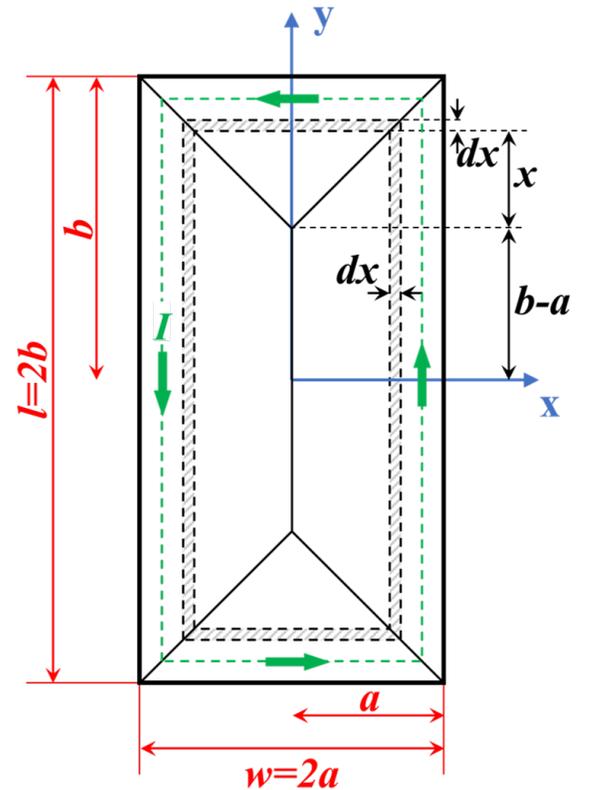

**Figure 1** Schematics for the calculation of $J_{c,mag}$ in the simplest Bean model scenario. The field is perpendicular to the sample rectangular cross-section $w \times l = 2a \times 2b$ (with $a < b$) and the current circulates in rectangular loops with the same absolute value of $J_{c,mag}$ at all points in the cross-section.

The total moment $m$ generated by the supercurrent can be calculated by integration:

$$m = 4J_{c,mag}t \int_0^a (b-a+x)x\, dx = 4J_{c,mag}t \left[\frac{a^3}{3} + \frac{a^2}{2}(b-a)\right]$$

Since the sample volume is $V = 4abt$, and the magnetization is by definition $M = m/V$, we obtain:

$$M = \frac{J_{c,mag}a}{2}\left(1 - \frac{a}{3b}\right).$$

The amplitude of the hysteresis loop is $\Delta M = 2M$, so:

$$\Delta M = J_{c,mag}a\left(1 - \frac{a}{3b}\right)$$

Substituting $a = w/2$ and $b = l/2$, we obtain the SI formula for $J_{c,mag}$:

$$J_{c,mag}[\tfrac{A}{m^2}] = \frac{2\Delta M[\tfrac{A}{m}]}{w[m]\left(1 - \frac{w[m]}{3l[m]}\right)} = \frac{2\Delta m[Am^2]}{V[m^3]w[m]\left(1 - \frac{w[m]}{3l[m]}\right)} \quad \text{(SI units)} \quad (1)$$

With $J_{c,mag}$ in [A/m²], $\Delta M$ in [A/m], $w$ and $l$ (being the full cross-section dimensions with $w < l$) in [m], and $V$ in [m³].

Since most instruments provide the moment in [emu], using eq. (1) requires first converting such [emu] moments to $m$ [Am²] with the correspondence 1 emu →10⁻³Am²[9], expressing $w$, $l$ and $t$ in [m] and $V$ in [m³]. If desired, the final conversion from A/m² to A/cm² is trivial.

We notice that the SI formula in eq.(1) is equivalent to the SI formula given by Talantsev and Tallon [2] and also matches Polichetti *et al.*'s eq.(3) in [4], confirming its validity. As reported by Talantsev and Tallon, an equivalent practical and frequently used formula using quantities in cgs-emu units on its right side is:

$$J_{c,mag}[\tfrac{A}{cm^2}] = 20\frac{Acm^2}{emu}\left(\frac{\Delta M[\tfrac{emu}{cm^3}]}{w[cm]\left(1 - \frac{w[cm]}{3l[cm]}\right)}\right) = 20\frac{Acm^2}{emu}\left(\frac{\Delta m[emu]}{V[cm^3]w[cm]\left(1 - \frac{w[cm]}{3l[cm]}\right)}\right) \quad \text{(converted for use with cgs-emu input data)} \quad (2)$$

with, $J_{c,mag}$ in A/cm², $\Delta M$ in [emu/cm³], $w$ and $l$ (being the full cross-section dimensions with $w < l$) in [cm], and $V$ in [cm³] and where $20\frac{Acm^2}{emu}$ is a conversion factor with dimensions, not just a pure number. Notice that eq.(2) is a practical converted formula that uses cgs-emu quantities as input data and an SI submultiple (A/cm²) as output, so it is neither a cgs-emu nor an SI formula: in fact, Ampere is not the unit of current in cgs-emu, but abA (abampere) or Bi (biot) (in cgs-esu and Gaussian units, the current is expressed in statA or esu/s). This hybrid unit formula has clear potential dangers.

**Comparison with Polichetti's formulae**

As verified above, Polichetti *et al.*'s eq.(3) in [4] correctly corresponds to SI eq.(1) here, but we also need to verify whether Polichetti *et al.*'s eq.(4) provides the correct $J_{c,mag}$ value. It must be noted that these expressions are not proper cgs-emu formulae (A, Ampere, is not a cgs-emu unit). We verify the equations in Table 1 by comparing the results of the calculation with the different formulae. In the table $\Delta m$, $w$, $l$ and $t$, $V$, $\Delta M$ are all quantities in SI units, as shown in the 2nd column. We use the already verified SI formula eq. (1) in the 2nd column as a reference to evaluate the other proposed equations. The quantities in the following columns indicate the conversion of $\Delta m$, $w$, $l$ and $t$, $V$, $\Delta M$ into the appropriate units requested in the formulae in the 1st row (e.g. in the 3rd column $\Delta m[emu] = (10^3 emu/(Am^2))\Delta m[Am^2]$), and we compare the results of the $J_{c,mag}$ calculation in the last row. From these calculations it is obvious that the formulae in the 2nd and 3rd columns, calculated from eqs. (1) and (2) of this paper, are correct with the expected factor of $10^{-4}$ comparing $J_{c,mag}$ in A/m$^2$ and in A/cm$^2$. However, columns 4 and 5 in Table 1, which use the 2 forms of Polichetti *et al.* eq.(4), are clearly both incorrect, one being 10 times too small and the other 10 times too big because the pre-factors are incorrect.

**Table 1** Here the quantities $\Delta m, w, l, t, V, \Delta M$ are all in SI units as showed in the 2$^{nd}$ column. In the following columns it is shown the conversion to obtain the quantities desired in the equations in the 1$^{st}$ row. E.g. in the 3$^{rd}$ column the value of $\Delta m[Am^2]$ in SI units has to be multiplied by $10^3 emu/(Am^2)$ to obtain the value in emu needed in the 1$^{st}$ raw formula; so $\Delta m[emu] = (10^3 emu/(Am^2))\Delta m[Am^2]$. Note that the units of the pre-factor 2 are not provided in ref. [4] for eq.(4) (1$^{st}$ form); to make the units match, we assume the pre-factor 2 is actually 2 Acm$^2$/emu (uncertainty marked by (?)).

| | Formulae resulting in **CORRECT** $J_{c,mag}$ values | | Formulae resulting into **INCORRECT** $J_{c,mag}$ values | |
|---|---|---|---|---|
| | Correct SI formula, eq.(1): $$J_{c,mag}\left[\frac{A}{m^2}\right] = \frac{2\Delta m[Am^2]}{V[m^3]w[m]\left(1-\frac{w[m]}{3l[m]}\right)}$$ | Converted hybrid cgs/SI formula, eq.(2): $$J_{c,mag}\left[\frac{A}{cm^2}\right] = 20\frac{Acm^2}{emu}\left(\frac{\Delta m[emu]}{V[cm^3]w[cm]\left(1-\frac{w[cm]}{3l[cm]}\right)}\right)$$ | Incorrect Polichetti *et al.*'s[4], eq.4, 1$^{st}$ form: $$\frac{2\times\Delta M\left[\frac{emu}{cm^3}\right]}{w[cm]\times\left(1-\frac{w[cm]}{3b[cm]}\right)} = J_{c,mag}[A/cm^2]$$ | Incorrect Polichetti *et al.*'s[4], eq.4, 2$^{nd}$ form: $$\frac{20\times\Delta M\left[\frac{A}{cm}\right]}{w[cm]\times\left(1-\frac{w[cm]}{3b[cm]}\right)} = J_{c,mag}[A/cm^2]$$ |
| $\Delta$moment | $\Delta m[Am^2]$ | $(10^3 emu/(Am^2))\Delta m[Am^2]$ | | |
| Small side | $w[m]$ | $(10^2 cm/m)w[m]$ | $(10^2 cm/m)w[m]$ | $(10^2 cm/m)w[m]$ |
| Large side | $l[m]$ (=b in eq.4 of [3]) | $(10^2 cm/m)l[m]$ | $(10^2 cm/m)l[m]$ ($l=b$ in eq.4 of [4]) | $(10^2 cm/m)l[m]$ ($l=b$ in eq.4 of [4]) |
| Thickness | $t[m]$ | $(10^2 cm/m)t[m]$ | | $10^2 t$ |
| Volume | $V[m^3] = wlt[m^3]$ | $(10^6 cm^3/m^3)V[m^3]$ | | |
| $\Delta$Magnetization | $\Delta M[A/m]$ $= \Delta m[Am^2]/V[m^3]$ | | $(10^{-3}emu/cm^3 \cdot m/A)\Delta M[A/m]$ | $(10^{-2}m/cm)\Delta M[A/m]$ |
| $J_{c,mag}$ results | $\frac{2\Delta m[Am^2]}{V[m^3]w[m]\left(1-\frac{w}{3l}\right)}$ Result in $[A/m^2]$ | $\frac{2\Delta m[Am^2]}{V[m^3]w[m]\left(1-\frac{w}{3l}\right)}10^{-4}\frac{m^2}{cm^2}$ Result in $[A/cm^2]$ | $\frac{2\Delta M[A/m]}{w[m]\left(1-\frac{w}{3l}\right)}10^{-5}\frac{m^2}{cm^2}(?) = \frac{2\Delta m[Am^2]}{V[m^3]w[m]\left(1-\frac{w}{3l}\right)}10^{-5}\frac{m^2}{cm^2}(?)$ Result in $[A/cm^2](?)$ | $\frac{2\Delta M[A/m]}{w[m]\left(1-\frac{w}{3l}\right)}10^{-3}\frac{m^2}{cm^2} = \frac{2\Delta m[Am^2]}{V[m^3]w[m]\left(1-\frac{w}{3l}\right)}10^{-3}\frac{m^2}{cm^2}$ Result in $[A/cm^2]$ |

**Conclusions**

The best approach to calculate $J_{c,mag}$ from magnetic measurements is by using SI units and formulae throughout, making the cgs-emu to SI conversion of magnetic moment $m$ right at the start. The present case shows that if a converted equation is desirable, its validity must be verified against a known valid equation to avoid the conversion error seen in ref. [4]. It is indeed probably true as suggested by Polichetti *et al.* that many errors have been made and not perhaps noticed. But in this case the error was important because a world record $J_c$ was claimed.

In general, results reporting exceptional $J_c$ values should be independently verified and assessed, preferably by transport measurements. If, as in [1], only magnetic moment measurements were made, it is also highly advisable to provide the raw data, equation used to convert $m$ to $J_c$ and all necessary information (like sample sizes for $J_{c,mag}$ calculation) needed for a proper data analysis to allow others to verify the results. This is how Talantsev and Tallon were able to find the error in the published $J_c$ result. Comparison to similar samples could also have provided a useful skepticism in this case because films using BaZrO$_3$ nanorod-containing vortex pinning centers were commercialized more than a decade ago by SuperPower and many reports have been published (e.g. [10], [11]). As an essential task of evaluating the many km of tape delivered for the 32 T user magnet at the NHMFL, it was observed that there was considerable variability of performance in such films. Francis *et al.* [12] evaluated the $J_c$ spread of a variety of samples and found that the volume%, diameter and spacing of the BaZrO$_3$ nanorods dominated the 4.2 K $J_c$ values. One of these samples, SP 215, has a remarkably similar BaZrO$_3$ distribution to the "record" sample of ref. [1]. The reported transport $J_c$ values of Francis are $J_c$(20 K, 7 T)~ 6.2 MA/cm$^2$ and $J_c$(4.2 K,8 T)~ 10.3 MA/cm$^2$[12], whereas Goyal's original estimations for the BZO sample were $J_{c,mag}$(20 K, 7 T)~ 60 MA/cm$^2$ and $J_{c,mag}$(4.2 K,7 T)~ 90 MA/cm$^2$[1]. The huge difference in $J_c$ between Goyal's and Francis' samples should have raised suspicions. Their $J_c$ values are indeed about a factor of 9-10 apart, due to the error caused by the incorrect conversion.

Finally, as advised more than a century ago by Giorgi [8] and pointed out by Goldfarb [7] in more recent years, it is probably time to abandon Gaussian or cgs-emu units for the preferable and better understood SI units and their immediate multiples and submultiples. We need to advocate now with the commercial magnetometer manufacturers!

**Acknowledgement**

This work was performed at the National High Magnetic Field Laboratory, which is supported by National Science Foundation Cooperative Agreements DMR-2128556, and by the State of Florida. Evaluation of REBCO coated conductors is supported by the DOE-Office of Fusion Energy grant DE-SC022011) and by a sub-award from ARPA-E through High Temperature Superconductor Inc.